%Paper: hep-ph/9404317
%From: YVONNE@URHEP.PAS.ROCHESTER.EDU
%Date: Thu, 21 Apr 1994 13:02:43 -0500 (EST)

\magnification=1200
\baselineskip=12pt
\tolerance=100000
\overfullrule=0pt
\rightline{UR-1352$\ \ \ \ \ \ \ $}
\rightline{ER-40685-803}

\bigskip
\baselineskip=18pt
\centerline{\bf CHIRAL INVARIANCE OF MASSIVE FERMIONS}

\bigskip

\bigskip

\centerline{Ashok Das}
\centerline{and}
\centerline{Marcelo Hott$^\dagger$}
\smallskip
\centerline{Department of Physics}
\centerline{University of Rochester}
\centerline{Rochester, NY l4627}

\bigskip
\bigskip
\bigskip

\centerline{\bf \underbar{Abstract}}

\medskip

We show that a massive fermion theory, while not invariant under the
 conventional chiral transformation, is invariant under a $m$-deformed
chiral transformation.  These transformations and the associated conserved
charges are nonlocal but reduce to the usual transformations and charges
when $m=0$.  The $m$-deformed charges commute with helicity and satisfy the
conventional chiral algebra.

\vskip 2.5in

\noindent $^\dagger$On leave
 of absence from UNESP - Campus de
Guaratinguet\'a, P.O. Box 205, CEP : 12.500, Guaratinguet\'a, S.P., Brazil
\vfill
\eject

\noindent {\bf I. \underbar{Introduction}:}

\medskip

Chiral invariance has played a fundamental role in the development of weak
interactions [1].  As we know, the Lagrangian for a massless fermion is
invariant under chiral transformations and since neutrino
is believed to be massless, the requirement of
 chiral invariance for the two component
neutrino fields leads to a V-A
structure of the weak interactions [2].
Chiral symmetry has also played a prominent role in understanding the low
energy properties of strong interactions through current algebra [3] and
PCAC [4].
 Chiral invariance of massive fermions
has also been studied in the literature [5] where it has been noted that the
Lagrangian for a massive fermion is invariant under the simultaneous
transformations
$$\eqalign{\psi &\rightarrow \gamma_5 \psi\cr
m &\rightarrow -m\cr}\eqno(1.1)$$
This transformation, also known as the ``mass reversal" transformation, is
not widely accepted and it is believed that chiral invariance is a property
of massless fermions.

It is known [6] that an interacting fermion theory at high temperature develops
a temperature dependent fermion mass where the mass grows with temperature.
 Thus, it would appear [7] that a massless, chiral invariant theory would have
its chiral symmetry broken by the temperature dependent mass.  The
understanding of this question becomes particularly relevant in connection
with the study of chiral symmetry restoration at high temperatures where
one conventionally believes that the dynamically broken chiral symmetry in
QCD is restored beyond a critical temperature.  As a first attempt to
understand this apparent conflict, we study the chiral invariance for
massive fermions at the level of the Lagrangian and the Hamiltonian.

Chiral invariance is a peculiar internal symmetry which does not allow a
fermion mass term in the Lagrangian.  The algebra of chiral symmetry, on
the other hand, imposes no restriction on mass.  In fact, being an
internal symmetry, $Q_5$ -- the generator of chiral transformation --
commutes with the energy-momentum $P_\mu$.
$$\left[ Q_5 , P_\mu \right] = 0 \eqno(1.2)$$
Therefore,
$$\left[ Q_5 , P^2 \right] = 0 \eqno(1.3)$$
and chiral invariance would appear to impose no condition on the value of
mass (which is given by the eigenvalue of the Casimir $P^2 = P_\mu P^\mu$).
 In fact, in section II, we will show that a massive fermion theory is
invariant under  $m$-deformed chiral transformations.  These are nonlocal
transformations which in the limit $m=0$ reduce to the conventional chiral
transformations.  In section III, we show that the conserved charges
associated with these $m$-deformed transformations satisfy the usual chiral
algebra.  We present our conclusions in section IV.

\medskip

\noindent {\bf II. \underbar{Chiral Invariance}:}

\medskip

To appreciate the generalized chiral invariance associated with massive
fermions, let us start with a free, massive fermion theory described by
$${\cal L} = \overline \psi \left( i \gamma^\mu \partial_\mu - m \right)
\psi = \overline \psi \left( i \gamma^0 \partial_0 + i \overrightarrow
 \gamma \cdot
 \overrightarrow \nabla - m \right) \psi \eqno(2.1)$$
We use the metric $\eta^{\mu \nu} = (+,-,-,-)$ and in our convention
$\gamma^{0^\dagger} = \gamma^0$, $ \overrightarrow
 \gamma^\dagger = -  \overrightarrow \gamma$ and
$\gamma_5 = i \gamma^0 \gamma^1 \gamma^2 \gamma^3$.  This Lagrangian
density is not invariant under a conventional chiral transformation.
However, we note that in the ultra relativistic limit, the fermion mass
becomes negligible and a chiral invariance would result.  Keeping this in
mind, we define a new field variable through a generalized Foldy-Wouthuysen
transformation [8] (more accurately a Cini-Touschek transformation) as
$$\psi (x) = U ( \overrightarrow \nabla ) \psi^\prime (x) \eqno(2.2)$$
where
$$\eqalign{U ( \overrightarrow \nabla ) = &{1 \over [ 2 ( m^2 -
 \overrightarrow \nabla^2 )^{
1/2} ( ( m^2 -  \overrightarrow \nabla^2 )^{1/2} + (
-  \overrightarrow \nabla^2 )^{1/2} ) ]^{1/2}}\cr
\noalign{\vskip 5pt}%
&\times \bigg( \big( m^2 -  \overrightarrow
 \nabla^2 \big)^{1/2} + \big( -  \overrightarrow \nabla^2
\big)^{1/2} -
i m \ { \overrightarrow \gamma \cdot
 \overrightarrow \nabla \over \big( -  \overrightarrow \nabla^2 \big)^{
1/2}} \bigg)\cr} \eqno(2.3)$$
and the derivatives are all acting to the right.  We note here that
conventionally such a transformation is written in momentum space as
$$U = e^{{1 \over 2m} \  \overrightarrow \gamma \cdot
 \overrightarrow p {\textstyle\theta} \ \left( {p
 \over m} \right)} \eqno(2.4)$$
with
$$\tan \left( {p \theta \over m}\right) = {m \over p} \qquad\qquad\qquad
 (p = |
 \overrightarrow p
|) \eqno(2.5)$$
Equivalently, one can also write
$$U = e^{{1 \over 2p}\  \overrightarrow \gamma
 \cdot  \overrightarrow p {\textstyle \theta} \left( {p \over m}
\right)} \eqno(2.6)$$
with
$$\tan \theta = {m \over p} \eqno(2.7)$$
However, we choose to work in the coordinate space and one can check
results when necessary by going over to the momentum space.

Under the field redefinition of Eq. (2.3), the Lagrangian density of Eq.
(2.1) takes the form (up to surface terms)
$$\eqalignno{{\cal L} &= \overline \psi \big( i \gamma^0 \partial_0
+ i  \overrightarrow \gamma \cdot
 \overrightarrow \nabla - m \big) \psi\cr
&= \overline \psi^\prime \bigg( i \gamma^0 \partial_0 + i
 \overrightarrow \gamma
\cdot  \overrightarrow \nabla\  {(m^2 -
 \overrightarrow \nabla^2 )^{1/2} \over
(-  \overrightarrow \nabla^2 )^{1/2}} \bigg) \psi^\prime\cr
&= \overline \psi^\prime \big( i \gamma^0 \partial_0 + i
 \overrightarrow \gamma \cdot
 \overrightarrow \nabla \hat O \big) \psi^\prime &(2.8)\cr}$$
where we have defined
$$\hat O = {(m^2 -  \overrightarrow \nabla^2 )^{1/2} \over
(-  \overrightarrow \nabla^2 )^{1/2}} \eqno(2.9)$$
We note that under the field redefinition, the Lagrangian has become
nonlocal, but it is now manifestly chiral invariant.  Before proceeding, we
note that the equations of motion in terms of the primed and unprimed
variables take the form
$$\eqalign{&\big( i \gamma^0 \partial_0 + i  \overrightarrow
 \gamma \cdot  \overrightarrow \nabla -
m \big)\psi = 0\cr
&\overline \psi \big( i \gamma^0 \overleftarrow \partial_0 + i
 \overrightarrow \gamma \cdot  \overleftarrow \nabla + m \big) = 0\cr}
\eqno(2.10)$$
$$\eqalign{&\big( i \gamma^0 \partial_0 + i  \overrightarrow
 \gamma \cdot  \overrightarrow \nabla \hat O\big) \psi^\prime = 0\cr
&\overline \psi^\prime \big( i \gamma^0 \overleftarrow \partial_0 + i
 \overrightarrow \gamma \cdot  \overleftarrow \nabla
 \overleftarrow{\hat O} \big) = 0\cr}
\eqno(2.11)$$
where $\overleftarrow{\hat O}$ stands for derivatives acting to the left.

Although there is no mass term in Eq. (2.11), they really describe massive
particles.  This is easily seen in momentum space where the equation has
the form
$$\eqalign{&\bigg( \gamma^0 p^0 - \overrightarrow \gamma \cdot
  \overrightarrow p \ {(m^2 + p^2 )^{1/2} \over p}\bigg) \psi^\prime = 0\cr
\noalign{\vskip 4pt}%
{\rm or,}\qquad &\bigg( \gamma^0 p^0 - \overrightarrow \gamma \cdot
  \overrightarrow p \ {(m^2 + p^2 )^{1/2} \over p}\bigg)^2 \psi^\prime = 0\cr
\noalign{\vskip 4pt}%
{\rm or,} \qquad&\big( p^{0^2}  - ( m^2 + p^2) \big)
 \psi^\prime = 0\cr
{\rm or,}\qquad &p^0 = \pm \big( m^2 + p^2 \big)^{1/2} \cr}\eqno(2.12)$$
On the other hand, the Lagrangian
 of Eq. (2.8) is invariant under the global chiral
transformations
$$\eqalign{\psi^\prime &\rightarrow e^{i \alpha \gamma_5} \psi^\prime\cr
\overline \psi^\prime &\rightarrow \overline \psi^\prime e^{i \alpha
\gamma_5}\cr}\eqno(2.13)$$
To understand the symmetry further, we need to construct the charges and
study their algebra.  To this end, we note that because the Lagrangian in
Eq. (2.8) is nonlocal, the derivation of the Noether current has to be
carried out carefully.  In fact, since the nontrivial operator
 $\hat O$ is not linear, namely, since
$$\hat O (AB) \not= (\hat OA)B + A(\hat O B) \eqno(2.14)$$
the continuity equation under an infinitesimal symmetry transformation for
the present case can be shown to be
$$\eqalign{\partial_0 \bigg( {\partial {\cal L} \over \partial \partial_0
\psi^\prime (x)}\ \delta_\epsilon \psi^\prime (x) \bigg) &+
\overrightarrow \nabla \cdot \bigg({\partial {\cal L} \over
\partial \overrightarrow \nabla \hat O \psi^\prime (x)}\
\delta_\epsilon \psi^\prime (x) \bigg)\cr
& = \bigg(
\overrightarrow \nabla \cdot {\partial {\cal L} \over
\partial \overrightarrow \nabla \hat O \psi^\prime (x)}
\bigg) \buildrel \leftrightarrow \over {\hat O} \delta_\epsilon \psi^\prime (x)
\cr}\eqno(2.15)$$
Here $ \buildrel \leftrightarrow \over {\hat O}
 = \hat O - \overleftarrow{\hat O}$.  For
the infinitesimal chiral transformations in Eq. (2.13) then, we can define
$$\eqalign{J^{\prime 0}_5 &= \overline \psi^\prime (x) \gamma_5 \gamma^0
\psi^\prime (x)\cr
\noalign{\vskip 4pt}
\overrightarrow J^\prime_5 &= \overline \psi^\prime (x) \gamma_5
\overrightarrow \gamma \hat O
\psi^\prime (x)\cr
\noalign{\vskip 4pt}%
J^\prime_5 &= \overline \psi^\prime (x) \gamma_5 \overrightarrow \gamma
\cdot \overleftarrow \nabla  \buildrel \leftrightarrow \over {\hat O}
\psi^\prime (x)\cr}\eqno(2.16)$$
That these operators satisfy the continuity equation (see Eq. (2.15))
$$\partial_0 J^{\prime 0}_5 + \overrightarrow \nabla \cdot
\overrightarrow J^\prime_5 = J^\prime_5 \eqno(2.17)$$
can be easily checked using the equations of motion in Eq. (2.11).  The
noncovariant nature of the currents is understandable since the field
redefinition in Eq. (2.2) or (2.4) or (2.6) is manifestly noncovariant.
However, the inhomogeneous term in Eq. (2.17) makes one wonder whether the
charge will be time independent.  We note from the structure of $J^\prime_5
(x)$ that
$$\int d^3x\ J^\prime_5 (x) = 0 \eqno(2.18)$$
(This can be seen more easily in momentum space if the coordinate space
derivation appears formal.)  Therefore, $J^\prime_5 (x)$ can really be
expressed as a total divergence although the expression is not simple.
Therefore, we continue to use the form of the continuity equation in
(2.17) and simply note that because of Eq. (2.18), the charge
$$Q^\prime_5 = \int d^3x\ \overline \psi^\prime (x) \gamma_5 \gamma^0
\psi^\prime (x) \eqno(2.19)$$
is a constant of motion.

The invariance of the action in terms of the original variables can now be
easily constructed.  Under
$$\eqalign{\psi (x) \rightarrow &U(\overrightarrow \nabla ) e^{i \alpha
\gamma_5} U (- \overrightarrow \nabla ) \psi (x)\cr
&= \bigg( \cos \alpha + i \gamma_5 \sin \alpha\  {(- \overrightarrow \nabla^2
)^{1/2} \over (m^2 - \overrightarrow \nabla^2 )^{1/2}} \
\bigg( 1 - i m\  {\overrightarrow \gamma \cdot \overrightarrow \nabla \over
\overrightarrow \nabla^2}\bigg) \bigg) \psi (x)\cr} \eqno(2.20)$$
it can be easily checked that the action associated with Eq. (2.1) is
invariant.  This is a manifestly nonlocal generalization of the
conventional chiral transformation and we note that when $m=0$ this reduces
to the usual chiral transformation associated with a massless field.  We
call these the $m$-deformed chiral transformations and we note that the
massive fermion theory of Eq. (2.1) is invariant under such
transformations.

The conserved currents associated with these transformations can now be
derived.  However, because the transformations are nonlocal, the derivation
has to be carried out with care.  Alternately, we can obtain the currents from
Eq. (2.17) through the inverse field redefinition.  In fact, it is
easy to check using Eq. (2.10) that
$$\eqalign{J^0_5 &= \overline \psi(x) U(\overleftarrow \nabla ) \gamma_5
\gamma^0 U(- \overrightarrow \nabla ) \psi (x)\cr
\overrightarrow J_5 &= \overline \psi(x) U(\overleftarrow \nabla ) \gamma_5
\overrightarrow \gamma \hat O
 U(- \overrightarrow \nabla ) \psi (x)\cr
J_5 &= \overline \psi(x) U(\overleftarrow \nabla ) \gamma_5
\overrightarrow \gamma \cdot   \overleftarrow \nabla
  \buildrel \leftrightarrow \over {\hat O} U(-
\overrightarrow \nabla ) \psi (x)\cr}\eqno(2.21)$$
satisfy the continuity equation
$$\partial_0 J^0_5 + \overrightarrow \nabla \cdot \overrightarrow J_5
= J_5 \eqno(2.22)$$
The charge
$$\eqalign{Q_5 &= \int d^3x\ J^0_5 (x)\cr
&= \int d^3x\ \overline \psi (x) \gamma_5 \gamma^0
{(- \overrightarrow \nabla^2 )^{1/2} \over (m^2 - \overrightarrow
\nabla^2)^{1/2}} \ \bigg( 1 - i m \ {\overrightarrow \gamma \cdot
\overrightarrow \nabla \over \overrightarrow \nabla^2} \bigg)
\psi (x) \cr}\eqno(2.23)$$
can be seen to reduce to the usual chiral charge when $m=0$ and, therefore,
can be thought of as a $m$-deformed chiral charge.  It can also be easily
seen to be a constant of motion in a variety of ways.  In fact, from Eq.
(2.20) we note that the first quantized generator of symmetry can be
identified with
$$Q_5 = \gamma_5 \left( {(-\overrightarrow \nabla^2 )^{1/2} \over
(m^2 - \overrightarrow \nabla^2 )^{1/2}} + i m
\ {\overrightarrow \gamma \cdot \overrightarrow \nabla \over
(m^2 - \overrightarrow \nabla^2)^{1/2} (- \overrightarrow \nabla^2 )^{1/2}}
\right) \eqno(2.24)$$
It is straightforward to check that this commutes with the first quantized
Hamiltonian
$$
\eqalignno{&H = -i \gamma^0 \overrightarrow \gamma \cdot \overrightarrow
\nabla + m \gamma^0 &(2.25)\cr
&[Q_5 , H ] = 0 &(2.26)\cr}$$
Let us also note that the helicity in the first quantized theory is defined
to be
$$h = - i \gamma_5 \gamma^0 {\overrightarrow \gamma \cdot \overrightarrow
\nabla \over (- \overrightarrow \nabla^2 )^{1/2}} \eqno(2.27)$$
It follows then that
$$[ Q_5 , h ] = 0 \eqno(2.28)$$
Namely, the $m$-deformed chiral charge continues to commute with helicity.
However, we note that
$$\left[ Q_5 , \gamma_5 \right] = -2 im {\overrightarrow \gamma
\cdot \overrightarrow \nabla \over
(m^2 - \overrightarrow \nabla^2 )^{1/2}
(- \overrightarrow \nabla^2 )^{1/2}} \eqno(2.29)$$
and the commutator vanishes for $m=0$.
\medskip

\noindent {\bf III. \underbar{Algebra of Charges}:}

\medskip

To study the algebra of the $m$-deformed chiral charges, let us consider,
for simplicity, a free, massive fermion theory where the fermion fields
belong to the fundamental representation of SU(2).
$${\cal L} = \overline \psi \left( i \rlap\slash{\partial} -m \right) \psi
= \overline \psi \left( i \gamma^0 \partial_0 + i \overrightarrow \gamma
\cdot \overrightarrow \nabla - m \right) \psi \eqno(3.1)$$
Once again, if we redefine the field variables as in Eqs. (2.2) and (2.3),
the Lagrangian will have the form
$${\cal L} = \overline \psi^\prime \left( i \gamma^0 \partial_0 + i
\overrightarrow \gamma \cdot \overrightarrow \nabla \
{(m^2 - \overrightarrow \nabla^2)^{1/2} \over
(- \overrightarrow \nabla^2 )^{1/2}} \right) \psi^\prime \eqno(3.2)$$
where $\psi^\prime (x)$ is a doublet of SU(2).  This Lagrangian is
invariant under global SU$_{\rm V}$(2) and SU$_{\rm A}$(2) transformations
defined by
$$\eqalign{\psi^\prime (x) &\rightarrow e^{{i \over 2}\ \overrightarrow
\alpha \cdot \overrightarrow \tau} \psi^\prime (x)\cr
\psi^\prime (x) &\rightarrow e^{{i \over 2}\ {\textstyle\gamma_5}
 \overrightarrow
\beta \cdot \overrightarrow \tau} \psi^\prime (x)\cr}\eqno(3.3)$$
where $\tau^a$'s represent the Pauli matrices -- the generators of SU(2).
Following the discussions of the previous section, we can construct the
conserved charges associated with these transformations and they are given
by

\vfill\eject

$$Q^{\prime a} = \int d^3x\ \overline \psi^\prime (x) \gamma^0
\big( {\tau^a \over 2}\big) \psi^\prime (x)$$
\line{\hfill $a=1,2,3\qquad$(3.4)}
$$Q^{\prime a}_5 = \int d^3x\ \overline \psi^\prime (x) \gamma_5
\gamma^0 \big( {\tau^a \over 2} \big) \psi^\prime (x)$$
Using the canonical anticommutation relations following from Eq. (3.2), it
can be easily checked that these charges satisfy the conventional chiral
algebra [9]
$$\eqalign{\big[ Q^{\prime a}, Q^{\prime b} \big] &= i\ \epsilon^{abc}
Q^{\prime c}\cr
\big[ Q^{\prime a}, Q^{\prime b}_5 \big] &= i\ \epsilon^{abc}
Q^{\prime c}_5\cr
\big[ Q^{\prime a}_5, Q^{\prime b}_5 \big] &= i\ \epsilon^{abc}
Q^{\prime c}\cr}\eqno(3.5)$$

The Lagrangian of Eq. (3.1) is, of course, invariant under global SU$_{\rm
V}$(2) transformations.  But it is also invariant under the $m$-deformed
SU$_{\rm A}$(2) transformations
$$\psi (x) \rightarrow \left( \cos {\beta \over 2} + i \
\gamma_5 \hat \beta \cdot \overrightarrow \tau \sin
{\beta \over 2} \ {(- \overrightarrow \nabla^2 )^{1/2} \over
(m^2 - \overrightarrow \nabla^2 )^{1/2}}\
\left( 1 - i m {\overrightarrow \gamma \cdot \overrightarrow \nabla
\over \overrightarrow \nabla^2} \right) \right) \psi (x) \eqno(3.6)$$
The conserved charges associated with these transformations can again be
constructed and have the form
$$\eqalign{Q^a &= \int d^3x \ \overline \psi (x) \gamma^0 \big(
{\tau^a \over 2}\big) \psi (x)\cr
Q^a_5 &= \int d^3x\ \overline \psi (x) \gamma_5 \gamma^0
\big( {\tau^a \over 2}\big) {(- \overrightarrow \nabla^2)^{1/2} \over
(m^2 - \overrightarrow \nabla^2 )^{1/2}} \
\bigg( 1 - i m {\overrightarrow \gamma \cdot \overrightarrow \nabla \over
\overrightarrow \nabla^2} \bigg) \psi (x) \cr}\eqno(3.7)$$
It is now straightforward to check, from the fundamental anticommutation
relations following from Eq. (3.1), that
$$\eqalign{\big[ Q^a, Q^b \big] &= i\ \epsilon^{abc}
Q^c\cr
\big[ Q^a, Q^b_5 \big] &= i\ \epsilon^{abc}
Q^c_5\cr
\big[ Q^a_5, Q^b_5 \big] &= i\ \epsilon^{abc}
Q^c\cr}\eqno(3.8)$$
Namely, the algebra of
 the $m$-deformed generators continues to be the usual
chiral algebra.

Finally, let us note that the operators
$$\eqalign{S^\prime &= {1 \over 2}\ \int d^3x\ \overline \psi^\prime (x)
\psi^\prime (x)\cr
\noalign{\vskip 4pt}
P^\prime &= - {i \over 2}\ \int d^3x\ \overline \psi^\prime (x)
\gamma_5
\psi^\prime (x)\cr
\noalign{\vskip 4pt}
\tilde Q^\prime_5 &= {1 \over 2}\ Q^\prime_5
= {1 \over 2}\ \int d^3x\ \overline \psi^\prime (x)
\gamma_5 \gamma^0
\psi^\prime (x)\cr}\eqno(3.9)$$

\line{satisfy the equal time SU(2) algebra \hfill}

$$\eqalign{\big[ S^\prime , P^\prime \big] &= i\ \tilde Q^\prime_5\cr
\big[ S^\prime , \tilde Q^\prime_5 \big] &= - i \ P^\prime\cr
\big[ P^\prime , \tilde Q^\prime_5 \big] &= i\ S^\prime\cr}\eqno(3.10)$$
The last two relations of Eq. (3.10), of course, represent the changes in
$S^\prime$ and $P^\prime$ under a chiral transformation and we recognize
the closed, equal time algebra to correspond to a SU(2) algebra [10].
Correspondingly, in terms of the original variables, the operators
$$\eqalign{S &= {1 \over 2}\ \int d^3x\ \overline \psi (x) \
{(- \overrightarrow \nabla^2 )^{1/2} \over
(m^2 - \overrightarrow \nabla^2 )^{1/2}} \
\bigg( 1 - i m {\overrightarrow \gamma \cdot \overrightarrow \nabla
\over \overrightarrow \nabla^2} \bigg) \psi(x)\cr
\noalign{\vskip 5pt}%
P &= - {i \over 2}\ \int d^3x\ \overline \psi (x) \gamma_5 \psi (x)\cr
\noalign{\vskip 5pt}%
\tilde Q_5 &= {1 \over 2}\ Q_5 = {1 \over 2}\ \int d^3x\ \overline \psi (x)
\gamma_5 \gamma^0\ {(-\overrightarrow \nabla^2 )^{1/2} \over
(m^2 - \overrightarrow \nabla^2 )^{1/2}}\
\bigg( 1 - i m \ {\overrightarrow \gamma \cdot \overrightarrow \nabla \over
\overrightarrow \nabla^2} \bigg) \psi (x)\cr}\eqno(3.11)$$

\noindent satisfy the equal time SU(2) algebra
$$\eqalign{[S,P] &= i\ \tilde Q_5\cr
\big[ S, \tilde Q_5 \big] &= -i\ P\cr
\big[ P, \tilde Q_5 \big] &= i\ S\cr}\eqno(3.12)$$

\medskip

\noindent {\bf IV. \underbar{Conclusion}:}

\medskip

Although a massive fermion theory is not invariant under a conventional,
global chiral transformation, we have shown that it is invariant under a
$m$-deformed chiral transformation.
 (This is quite similar to the fact that a spontaneously broken
 massive gauge theory has a $m$-deformed gauge invariance (transformations
depending on mass or vacuum expectation value) at the Lagrangian level.
  The difference is that
the $m$-deformed gauge transformations are local and further, since the
number of degrees of freedom is not the same for
$m=0$ and $m \not= 0$, in the case of a gauge theory, one needs
additional fields (namely, the Higgs field) for this.)
  These transformations as well as the
corresponding conserved charges are manifestly nonlocal but reduce to the
usual transformations
 and charges when $m=0$.  The $m$-deformed
charges continue to commute with the helicity operator and satisfy the
usual chiral algebra.  A massive fermion theory is also known to have a
second $\gamma_5$ invariance [11] where, however, the generators have explicit
time dependence [12].
 Our discussion so far has been at the level of operators, the Lagrangian
and the Hamiltonian.  The question of chiral symmetry restoration at high
temperature needs further careful study of the structure of the Hilbert
space and will be reported in a later publication.

We would like to thank Profs. V.S. Mathur and S. Okubo for discussions and
comments.  This work was supported in part by U.S. Department of Energy
Grant No. DE-FG-02-91ER40685.  One of us (M.H.) would like to thank the
Fundac\~o de Amparo a Pesquisa do Estado de S\=ao Paulo for the financial
support.

\vfill\eject

\noindent {\bf \underbar{References}:}

\medskip

\item{1.} R.E. Marshak, ``Conceptual Foundations of Modern Particle
Physics", World Scientific (1993).

\item{2.} R.E. Marshak and E.C.G. Sudarshan, Proc. Int. Conf.
Elem. Part., Padua (1957); Phys. Rev. {\bf 109}, 1860 (1958);
R.P. Feynman and M. Gell-Mann, Phys. Rev. {\bf 109}, 193 (1958);
 A. Salam, Nuovo Cimento {\bf 5}, 299 (1957); L. Landau, Nucl.
 Phys. {\bf 3}, 127 (1957).

\item{3.} M. Gell-Mann, Phys. Lett. {\bf 8}, 214
(1964).

\item{4.} S. Adler, ``Current Algebras and Applications to Particle
Physics", Benjamin (1968).

\item{5.} J. Tiomno, Nuovo Cimento {\bf 1}, 226 (1955);
 J.J. Sakurai, Nuovo Cimento {\bf 7}, 649 (1958).

\item{6.} H.A. Weldon, Phys. Rev. {\bf D26}, 2789 (1982); V.V. Klimov, Sov.
J. Nucl. Phys. {\bf 33}, 934 (1981).

\item{7.} L.N. Chang, N.P. Chang and K.C. Chou, Phys. Rev. {\bf D43}, 596
(1991); L.N. Chang and N.P. Chang, Phys. Rev. {\bf D45}, 2988 (1992); N.P.
Chang, ``Braaten-Pisarski Action $\dots$", CCNY-HEP-93-15; N.P. Chang,
``Chiral Morphing", CCNY-HEP-94-6.

\item{8.} L.L. Foldy and S. Wouthuysen, Phys. Rev. {\bf 78}, 29
(1950); S. Tani, Prog. Theor. Phys. {\bf 6}, 267 (1951); M. Cini and B.
Touschek, Nuovo Cimento {\bf 7}, 422 (1958).

\item{9.} B.W. Lee, ``Chiral Dynamics", Gordon and Breach (1972).

\item{10.} This is the analogue of the chirality algebra discussed in ref.
7.

\item{11.} J. Soto and R. Tzani, Phys. Lett. {\bf B297}, 358 (1992); J.
Soto and R. Tzani, Barcelona preprint UB-ECM-PF-93-1.

\item{12.} A. Das and V.S. Mathur, Phys. Rev. {\bf D49}, 2508 (1994).

\end